\documentclass[9pt,twocolumn,twoside]{osajnl}
\journal{ol} % Choose journal (ao, aop, josaa, josab, ol, pr)
\usepackage{graphicx}
\usepackage{lipsum}
\usepackage{bigints}
\usepackage{nicefrac}
\usepackage[capitalise]{cleveref}
\usepackage{bm}
\let\oldsqrt\sqrt
\def\sqrt{\mathpalette\DHLhksqrt}
\def\DHLhksqrt#1#2{%
\setbox0=\hbox{$#1\oldsqrt{#2\,}$}\dimen0=\ht0
\advance\dimen0-0.2\ht0
\setbox2=\hbox{\vrule height\ht0 depth -\dimen0}%
{\box0\lower0.4pt\box2}}
\setboolean{shortarticle}{true}
\title{Broadband Resonator-Waveguide Coupling for Efficient Extraction of Octave Spanning Microcombs}

\author[1,2,$\dagger$]{Gregory Moille}
\author[1,3]{Qing Li}
\author[4,5]{Travis C. Briles}
\author[4,5]{Su-Peng Yu}
\author[4,5]{Tara Drake}
\author[1,2]{Xiyuan Lu}
\author[1,2]{Ashutosh Rao}
\author[1]{Daron Westly}
\author[4,5]{Scott B. Papp}
\author[1,6,$\dagger$]{Kartik Srinivasan}

\affil[1]{Microsystems and Nanotechnology Division, National Institute of Standards and Technology, Gaithersburg, MD 20899, USA}
\affil[2]{Institute for Research in Electronics and Applied Physics and Maryland Nanocenter, University of Maryland,College Park, Maryland 20742, USA}
\affil[3]{Electrical and Computer Engineering, Carnegie Mellon University, Pittsburgh, PA 15213, USA}
\affil[4]{Time and Frequency Division, National Institute of Standards and Technology, 385 Broadway, Boulder, CO 80305, USA}
\affil[5]{Department  of  Physics,  University  of  Colorado,  Boulder,  Colorado,  80309,  USA}
\affil[6]{Joint Quantum Institute, NIST/University of Maryland, College Park, Maryland 20742, USA}
\affil[$\dagger$]{gregory.moille@nist.gov, kartik.srinivasan@nist.giv}

\begin{abstract}
Frequency combs spanning over an octave have been successfully demonstrated in Kerr nonlinear microresonators on-chip. These micro-combs rely on both engineered dispersion, to enable generation of frequency components across the octave, and on engineered coupling, to efficiently extract the generated light into an access waveguide while maintaining a close to critically-coupled pump. The latter is challenging as the spatial overlap between the access waveguide and the ring modes decays with frequency. This leads to strong coupling variation across the octave, with poor extraction at short wavelengths. Here, we investigate how a waveguide wrapped around a portion of the resonator, in a pulley scheme, can improve extraction of octave-spanning microcombs, in particular at short wavelengths. We use coupled mode theory to predict the performance of the pulley couplers, and demonstrate good agreement with experimental measurements. Using an optimal pulley coupling design, we demonstrate a 20~dB improvement in extraction at short wavelengths compared to straight waveguide coupling.
\end{abstract}

\colorlet{mycol}{red!70!lime}
\newcommand{\hg}[1]{#1}
\doi{\url{https://doi.org/10.1364/OL.44.004737}}
\setboolean{displaycopyright}{false}
\begin{document}

\maketitle
\begin{figure}[t]
    \centering
    \includegraphics{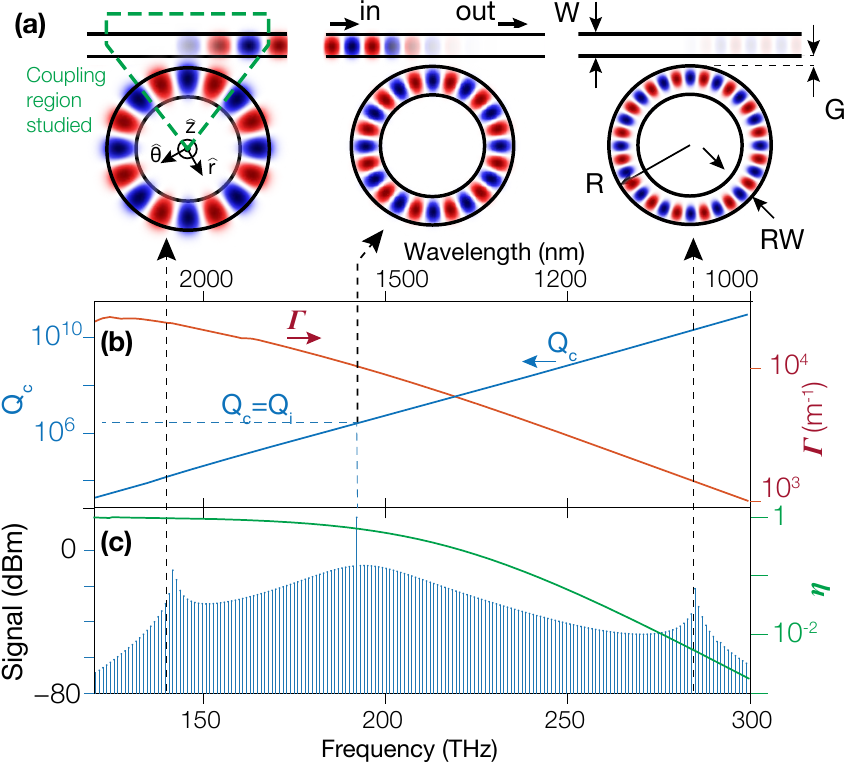}
    \caption{\label{fig:fig1} (a) Schematic of the optical modes of the ring resonator coupled to a straight waveguide at the DW and pump frequencies. (b) Coupling quality factor $Q_\mathrm{c}$ (blue, left y-axis) and ring/waveguide mode overlap $\Gamma$ at the coupling point (red, right y-axis), for $R=23.3$~$\mu$m, $RW=1600$~nm, $W=550$~nm, and $G=700$~nm. (c) \hg{Simulated intracavity comb spectrum. The variation in $Q_\mathrm{c}$ in (b) indicates that high frequency comb components will not be well extracted, as shown through the extraction efficiency $\eta$ assuming $Q_i=3\times10^6$ (green).}}
\end{figure}

Kerr solitons generated in nonlinear microresonators~\cite{kippenberg_dissipative_2018} are promising for many applications in telecommunications~\cite{Wu2018}, range measurement~\cite{Suh884}, and optical frequency metrology~\cite{Spencer:2018cd7}. However, for frequency metrology in particular, many applications require octave-spanning bandwidth for full stabilization through $f$-2$f$ technique~\cite{Udem2002}. Suitable engineering of the resonator dispersion profile for octave bandwidth~\cite{Okawachi:14,li2015octave}, or even super-octave bandwidth~\cite{Moille:2018ba}, has been widely reported and octave-spanning soliton frequency combs have been demonstrated~\cite{Li2017,Pfeiffer2017}, along with $f$-2$f$ stabilization~\cite{Spencer:2018cd7,Briles:20188c6}. Such stabilization requires sufficient power at the frequencies of interest. This ultimately depends not only on the generated intracavity field and ability to take advantage of effects like targeted dispersive wave (DW) emission~\cite{li2015octave,Brasch2016a,yi_single-mode_2017}, but also on the extraction of the intracavity field, usually through evanescent coupling to an in-plane waveguide (or waveguides) for microring resonators. Efficient extraction over an octave of bandwidth is particularly non-trivial due to the wavelength dependence of both the phase matching and the spatial mode overlap between the resonator and waveguide modes.\\
\indent In this letter, we characterize an approach to overcome this challenge, particularly at short wavelengths, based on a pulley configuration in which a portion of the access waveguide is wrapped around the microring~\cite{Hosseini2010}. Though utilized in our recent octave-spanning microcomb works~\cite{Li2017,Spencer:2018cd7,Briles:20188c6}, this approach was not studied in detail. Here, we present a basic coupled mode theory (CMT)
formalism to design the pulley to improve resonator-waveguide coupling, thus comb extraction at short wavelengths, while maintaining desirable coupling in the pump and long wavelength bands. One consequence of pulley coupling is the introduction of narrow spectral windows in which essentially no coupling occurs, due to complete phase-mismatch between the ring and waveguide modes. Consequently, it is important to control the spectral position of these windows in which no coupling occurs, which we refer to as anti-phase-matched frequencies, so that they are separated from the regions of interest, namely the pump and DW frequencies. Experimentally, we validate both the control of the pulley anti-phase-matched frequencies and the enhancement of short wavelength extraction, by $\approx$~20~dB relative to conventional point coupling using straight waveguides.\\
\indent A number of computational approaches have been used to model coupling between ring resonators and waveguides~\cite{chin_design_1998,Hosseini2010,bahadori_design_2018}. Here, we model resonator-waveguide coupling in an integrated planar geometry by considering only the region over which their fields interact (\cref{fig:fig1}(a)), with microring outer radius $R$ and ring width $RW$ separated by a gap $G$ from the coupling waveguide of width $W$. Using the spatial CMT formalism common to waveguide directional couplers~\cite{yariv1973, yariv2007photonics}, we determine the per-pass coupling coefficient from resonator to waveguide $\kappa_\mathrm{r \rightarrow wg}$, based on the overlap of the ring and waveguide fields, integrated over the coupling portion. From there, the coupling quality factor $Q_\mathrm{c}$ is obtained, and compared to a typical resonator intrinsic quality factor ($Q_\mathrm{i}$) to gauge whether the coupling is at an appropriate level, namely close to critical coupling at the pump $Q_\mathrm{c}\approx Q_\mathrm{i}$.\\
\indent The coupling coefficient between the ring and the waveguide is:
\begin{equation}
  \kappa_\mathrm{r \rightarrow wg}  = \bigintss_L  \Gamma(\omega, l) \mathrm{e}^{i \phi} \mathrm{d}l
  \label{eq:kappa}
\end{equation}
\noindent with $L$ being the optical path and $\Gamma$ the overlap of the ring mode projected onto the waveguide mode as:
\begin{equation}
    \Gamma(\omega, l) = \frac{i \omega}{4} \bigintss_S (\varepsilon_{\mathrm wg}- \varepsilon_{\mathrm R}) \bm{E_{\mathrm R}^ *}\cdot \bm{E_{\mathrm{wg}}} \; \mathrm{d}r\mathrm{d}z \label{eq:Gamma}
\end{equation}
with $\omega$ the angular frequency,  $\bm{E_{\mathrm{R, wg}}}$  the electric field of ring and waveguide mode respectively, normalized such that $P =\frac{1}{2} \iint \left(\bm{E}\times\bm{H^*}\right)\cdot\hat{\theta}\mathrm{d}r\mathrm{d}z=1$\hg{~\cite{yariv2007photonics}}, with $r$, $\theta$, and $z$ being the radial, azimuthal, and vertical directions as taken from the center of the ring (\cref{fig:fig1}(a)). $\varepsilon_{\mathrm R, wg}$ is the dielectric permittivity considering only the ring and waveguide, respectively. The accumulated phase term in \cref{eq:kappa} corresponds to
$\phi = l \sqrt{\left(\nicefrac{\Delta \beta}{2}\right)^2 + \Gamma^2}$, where \hg{$\Delta\beta = \frac{m}{R_\mathrm {wg}} - n_\mathrm {eff}^\mathrm {wg}\frac{\omega}{c_\mathrm{0}} = \frac{\omega}{c_\mathrm{0}}\Delta n_\mathrm{eff}^\mathrm{wg}$} is the difference of propagation constant between the ring and the waveguide mode, within the waveguide, with $m$ the azimuthal mode number of the ring for a given resonance frequency, $R_\mathrm{wg}=R+G+\nicefrac{W}{2}$, and $c_{0}$ the speed of light in vacuum.\\
\indent The effects of the coupling coefficient, $\kappa_\mathrm{r \rightarrow wg}$, the phase-mismatch $\Delta \beta$ and the mode overlap $\Gamma$ can be combined into a single quantity describing the coupling strength known as the coupling quality factor $Q_\mathrm{c}$, defined as:
\begin{equation}
    Q_{\mathrm c} = \omega\frac{n^\mathrm{R}_{\mathrm g}}{c_\mathrm{0}} \frac{2\pi R}{\left| \kappa_{\mathrm r \rightarrow wg}\right|^2}
    \label{eq:Qc}
\end{equation}
\noindent where $n^{\mathrm R}_{\mathrm g}$ is the group index of the ring resonator. It is convenient to compare $Q_\mathrm{c}$ with the intrisic quality factor $Q_\mathrm{i}$, and to define the extraction efficiency as $\eta = \left(1 + \nicefrac{Q_\mathrm{c}}{Q_\mathrm{i}}\right)^{-1}$.\\
\begin{figure}[t]
    \centering
    \includegraphics{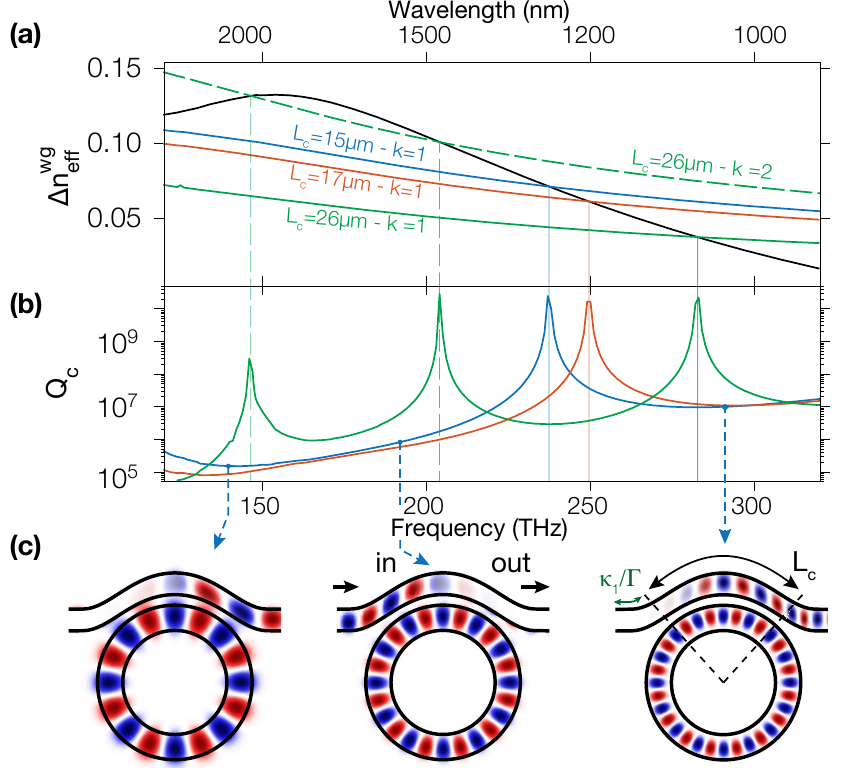}
    \caption{\label{fig:2}(a) Difference in effective index between the ring and waveguide mode $\Delta n_\mathrm{eff}^\mathrm{wg}$ (black) and the anti-phase matching condition (right hand side of \cref{eq:pulley_resonances}) for $L_\mathrm{c}$=\{15, 17, 26\}~$\mu$m (blue, red, green). $L_\mathrm{c}$=26~$\mu$m supports an anti-phase condition for both the first and second order (solid and dashed green lines). (b) $Q_{c}$ for the corresponding pulley lengths. (c) Cartoon depicting the behavior of the electric field for the three frequencies shown as dashed lines in (b), corresponding to the DW and pump frequencies, for $L_\mathrm{c}=$15$\mu$m}
\end{figure}
\begin{figure*}[!t]
    \centering
    \includegraphics{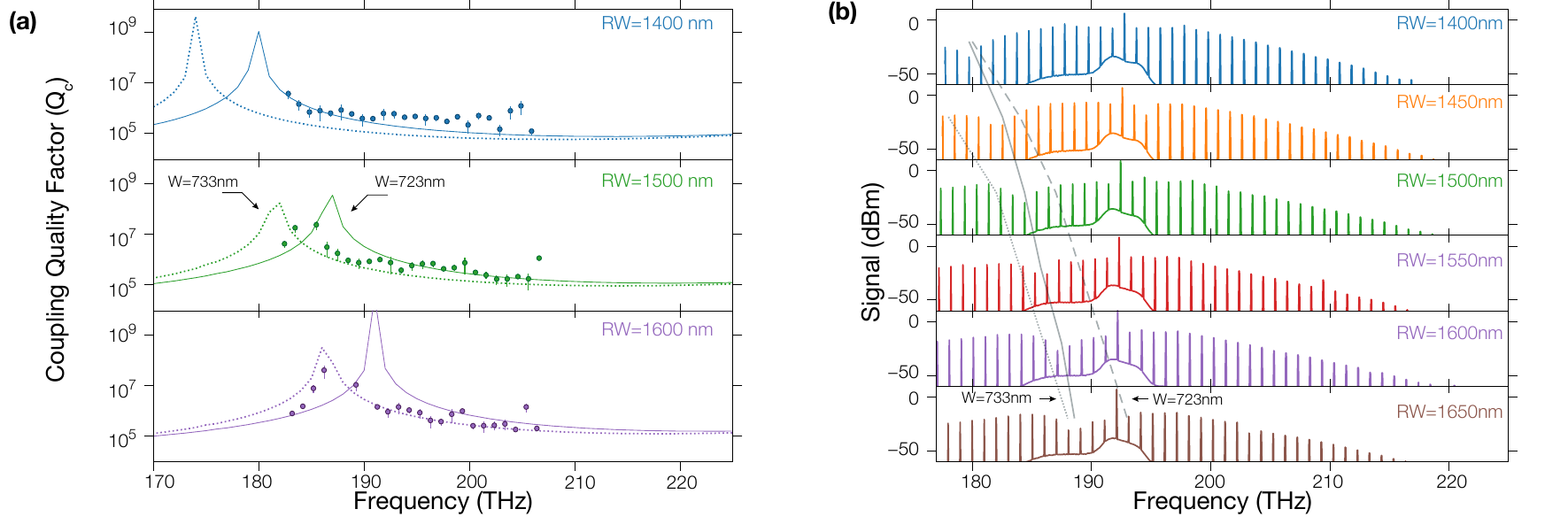}
    \caption{\label{fig:3}(a) Linear transmission measurement of $Q_{\textrm{c}}$ (circles) for a $L_\mathrm{c}=$ 40~$\mu$m pulley waveguide, for a ring width of 1400~nm (blue), 1500~nm (green) and 1600~nm (puple). The solid and dashed lines represent the simulated $Q_\mathrm{c}$ for a waveguide width of 723~nm and 733~nm respectively. \hg{The error bars we report are one standard deviation values based on nonlinear least squares fits to the model from Ref~\cite{borselli_beyond_2005}}. (b) MI combs obtains for different ring widths. The dips in the comb profiles correspond to the pulley anti-phase-matched frequencies and are highlighted through the solid lines. The dotted and dashed lines represent the theoretical position of the pulley anti-phase-matched frequencies, and hence the expected dip in the MI comb spectra, for a waveguide width of 723~nm and 733~nm respectively}
\end{figure*}
\indent The basic challenge that we address is conceptually illustrated in \cref{fig:fig1}, and is easy to explain using this CMT framework. Straight waveguide coupling to a ring resonator involves a limited interaction length over which the waveguide and resonator modes spatially overlap, leading to close to a point-like coupling region, particularly for small diameter rings. At long wavelengths (low frequencies) and for a carefully chosen gap size $G$, the overlap $\Gamma$ can be appreciable enough that a short interaction length is adequate, yielding $Q_{\textrm{c}}$ comparable to $Q_{\textrm i}$ (i.e., critical coupling). However, as seen in \cref{fig:fig1}(a,b), as the wavelength decreases (frequency increases), each mode is more confined, leading to a reduction in $\Gamma$, and $Q_{\textrm{c}}$ increases exponentially with frequency. This results in poor coupling at short wavelengths, with $Q_{\textrm{c}}$ orders of magnitude higher than $Q_{\textrm{i}}$ (\cref{fig:fig1}(b)). This is problematic for octave-spanning combs, as illustrated in \cref{fig:fig1}(c). Here, the spectrum of the intracavity field is simulated by solving the Lugiato-Lefever equation with the open-source \textit{pyLLE} package~\cite{Moille:2019291}, for a geometry appropriate for supporting octave-span operation. Though the dispersion has been engineered to support nearly harmonic dispersive waves at 280~THz and 155~THz, the $>$100~$\times$ difference in $Q_{\textrm{c}}$ will lead to very different out-coupled powers (given a $Q_{\textrm{i}}~\approx~3\times10^6$ that is not expected to significantly vary with wavelength), as seen in the plot of $\eta$ in Fig.~\ref{fig:fig1}(c). This will be a major impediment to direct
self-referencing. \\
\indent To overcome this issue, it is possible to increase the interaction length between the waveguide and the ring by wrapping the former around the latter, resulting in a pulley coupling design~\cite{Hosseini2010} shown schematically in Fig.~\ref{fig:2}(c). We note that the overlap coefficient $\Gamma$ in \cref{eq:Gamma} is independent of position along the optical path $L$, and the accumulated phase has to be accounted for across the pulley length $L_\mathrm{c}$, \hg{length for which the gap is constant between the ring and the waveguide}. Thus \cref{eq:kappa} becomes:
\begin{equation}
    \kappa_{\mathrm r \rightarrow wg}  = \Gamma (\omega) \bigintss_{{L_{\mathrm c}}} \mathrm{e}^{i \phi} \mathrm{d}L = \Gamma L_\mathrm{c}\mathrm{sinc}\left({L_{\mathrm c}}\sqrt{ \left(\nicefrac{ \Delta\beta}{2}\right)^2 + \Gamma^2}\right)
    \label{eq:kappasinc}
\end{equation}

\noindent with $\mathrm{sinc}(x) = \nicefrac{\mathrm{sin}(x)}{x}$. Hence \cref{eq:Qc} can be rewritten as:
\begin{equation}
  Q_\mathrm{c} = \omega \frac{n_\mathrm{g}^\mathrm{R}}{c_\mathrm{0}}  2\pi R \left[  \Gamma L_\mathrm{c}\mathrm{sinc}\left({L_{\mathrm c}}\sqrt{ \left(\nicefrac{ \Delta\beta}{2}\right)^2 + \Gamma^2}\right) \right]^{-2}
  \label{eq:Qcsinc}
\end{equation}%
\noindent The above only accounts for the region where the resonator-waveguide gap is constant, and not where the waveguide bends towards and away from the ring, as seen in \cref{fig:2}(c). To account for this, we evaluate \cref{eq:kappa} in these regions, resulting in a coupling coefficient $\kappa_1$. The ratio $\nicefrac{\kappa_1(\omega)}{\Gamma_\mathrm{pulley}(\omega)}$ gives the effective length of the curved portion (\cref{fig:2}(c)). Hence, one can then introduce an effective pulley length $\widetilde{L_\mathrm{c}}(\omega) = L_\mathrm{c} + 2 \nicefrac{\kappa_1(\omega)}{\Gamma_\mathrm{pulley}(\omega)}$ that replaces the pulley length in \cref{eq:kappasinc}.

Interestingly, \cref{eq:kappasinc,eq:Qcsinc} suggest that resonances in the coupling will happen according to:
\begin{equation}
\Delta n_\mathrm{eff}^\mathrm{wg}= 2\frac{c_{\mathrm 0}}{\omega} \sqrt{ \left(\nicefrac{k\pi}{\widetilde{L_\mathrm {c}}}\right)^2 - \Gamma^2 } \;\quad;\; k \in \mathbb N.
  \label{eq:pulley_resonances}
\end{equation}%
\noindent Physically, these resonances correspond to locations where the access waveguide and the ring waveguide are anti-phase-matched, and are not observed for a straight waveguide due to the limited interaction length (over which the gap is continuously varying). To investigate further, we calculate $Q_\mathrm{c}$ using parameters that correspond to the experimental system studied, that is, 780~nm thick silicon nitride (Si\textsubscript{3}N\textsubscript{4}) microrings that are symmetrically clad in silica (SiO\textsubscript{2}), with $R=$23.3~$\mu$m. We pick $RW=1600$~nm (resulting in the simulated frequency comb shown in \cref{fig:fig1}(b)), along with $W=550$~nm and $G=700$~nm. As shown in \cref{fig:2}, the anti-phase-matching condition results in sharp peaks in $Q_\mathrm{c}$ for frequencies that vary with $L_c$. At these frequencies, regardless of the overlap between the ring and the waveguide modes, no transfer of energy occurs. The behavior of $Q_\mathrm{c}$ on either side of the resonances is important for octave-spanning comb applications. On the blue side (short wavelengths), the overall increase in interaction length results in smaller $Q_\mathrm{c}$ (improved coupling) than in the straight waveguide case. On the red side (longer wavelengths), the difference in $\Delta n_\mathrm{eff}^\mathrm{wg}$, accumulated over $L_\mathrm{c}$, keeps $Q_\mathrm{c}$ larger than in the straight waveguide case, where the rings are generally overcoupled. The net result is a reduced wavelength-dependence in $Q_{\textrm{c}}$ (outside of the anti-phase-matched window) than for a straight waveguide. We also note that when the pulley is sufficiently long, higher orders of anti-phase-matching can be satisfied (\textit{i.e.}, $k > 1$), leading to multiple resonances in $Q_\mathrm{c}$. \\
\indent To validate the CMT modeling, we first verify the pulley resonance behavior through linear transmission measurements of devices designed to show a $Q_\mathrm{c}$ resonance within the $182$~THz to $207$~THz tuning range of our laser source. We keep the pulley parameters fixed, namely gap $G=800$~nm, waveguide width $W=750$~nm, and pulley coupling length $L_\mathrm{c}=40$~$\mu$m, while the ring width $RW$ is varied. This results in variation of the effective index of the microring $n_\mathrm{eff}^\mathrm{ring}$, leading to a modification of the anti-phase-matching condition and hence the spectral position of the corresponding frequencies. By fitting approximately 25 resonances of the first order transverse electric (TE) mode family that appear within the laser scan range, we extract the spectral dependence of $Q_\mathrm{c}$ (\cref{fig:3}(a)) for each $RW$, taking into account internal losses, coupling, and backscattering~\cite{borselli_beyond_2005}.\\
\begin{figure}[t]
    \centering
    \includegraphics{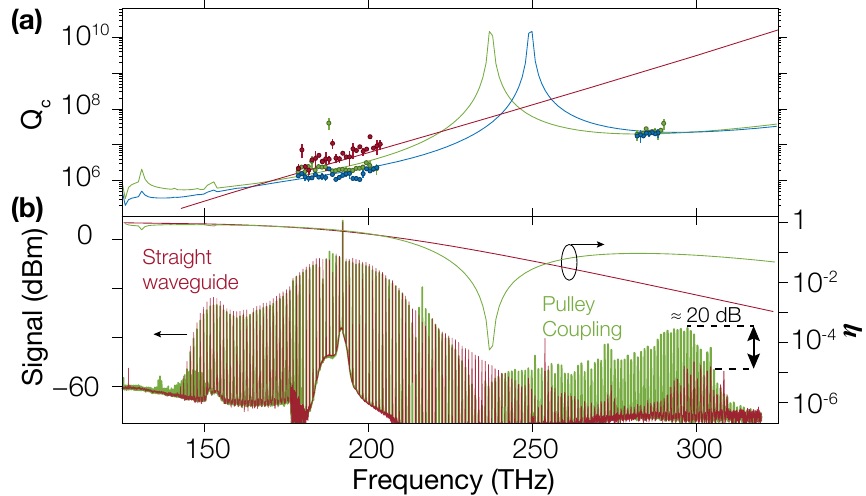}
    \caption{\label{fig:4}(a) Linear transmission measurement of $Q_{\textrm{c}}$ (circles) for a straight waveguide (red), and pulley coupling with $L_c$=15~$\mu$m (green) and $L_c$=17~$\mu$m (blue). Error bars are one standard deviation values extracted from nonlinear least squares fits. (b) MI comb spectra for straight waveguide (red) and pulley coupling with $L_c$=15~$\mu$m (green) with estimated extraction efficiency $\eta$ assuming $Q_{i}=3\times10^6$ (right axis).}
\end{figure}
\indent Simulations of $Q_{\textrm{c}}$ through CMT match both the values and the trend of the experimental $Q_{\textrm{c}}$, including the divergence at resonance. Moreover, one can reproduce the variation in $Q_{\textrm{c}}$ anti-phase-matching spectral position with $RW$ due to the change of effective index. The simulations also show the difference of sensitivity in dimension between the ring and the waveguide. As the waveguide is narrower with a mode less confined than the ring, a small variation of its width results in a significant change in its effective index. Hence, by only changing the waveguide width by 10~nm, the anti-phase matching frequency shifts by about 5~THz. To achieve the same shift, one needs to modify the $RW$ by 50~nm. This gives the ability to tune the position of the pulley anti-phase-matched frequency while keeping the microring resonator at a fixed geometrical dimension that is likely already dispersion-optimized.\\
\indent Outside of the tunable laser range, it is possible to extract the position of the pulley anti-phase-matched frequency by measuring the spectra of modulation instability (MI) combs generated through strong pumping ($P_\mathrm{pmp}=200$~mW) of the resonators at $\approx$~1550~nm, on the blue-detuned side of a cavity resonance. The spectral components in these MI combs are not phase-locked, and the overall comb acts as a quasi-continuous-wave, spectrally broadband source. Thus, the pulley coupling can be studied using these states over a spectral range as broad as the comb bandwidth. To confirm this, we measure the MI comb spectra of the devices characterized linearly (\cref{fig:3}(b)). We observe that the position of the anti-phase-matched frequency obtained through linear characterization of the device (by extracting $Q_{\textrm{c}}$) and through measuring the position of the dip in the MI comb are consistent. The latter method also agrees with the CMT simulations, and the position of the anti-phase-matched frequency is within the uncertainty in the fabricated geometry.\\
\indent We now compare a pulley coupling design optimized for extraction of an octave-spanning microcomb, namely, with a coupling anti-phase-matched frequency in-between the pump and the short wavelength DW, against straight waveguide coupling for the same ring parameters. We first characterized $Q_{\textrm{c}}$ (through linear transmission measurements) in both the pump band and near the short wavelength DW, around 193~THz and 280~THz, respectively (\cref{fig:4}(a)) \hg{using two continuous tunable laser centered around 1550 nm and 1050nm.} We were unable to measure any resonance of the first order TE mode in the 280~THz range for the straight waveguide, as expected from simulations where $Q_{\textrm{c}}$ is orders of magnitude higher than the expected $Q_{\textrm{i}}\approx 3\times 10^6$ (as measured in other bands). In contrast, the pulley devices, for both $L_c$=15~$\mu$m and $L_c$=17~$\mu$m, exhibit a difference in $Q_{\textrm{c}}$ of only one order of magnitude between the two bands, and show good agreement with the values predicted by the CMT.
Finally, from MI comb spectra (\cref{fig:4}(b)), the advantage of using the pulley coupling approach for extraction is apparent. Pumping both the straight waveguide and the $L_c$=15~$\mu$m pulley devices such that the long DW and overall comb shape are the same, the pulley coupling shows a clear advantage in extracting the short DW with $>20$~dB increase in power obtained. \hg{This enhancement of short DW extraction has recently been applied in studies of octave-spanning soliton microcombs~\cite{Spencer:2018cd7,Li2017,Briles:20188c6,SuPengPhysRevApplied2019}}.\\
\indent In conclusion, we have presented a CMT formalism to design pulley couplers to help with the extraction of octave-spanning spectra from chip-integrated, microring-based frequency combs.  We use the CMT to elucidate the roles of phase-mismatch and spatial overlap in the wavelength-dependent coupling spectrum.  Finally, we show that using such pulley coupling increases by $\approx$20~dB the extraction of the short wavelength part of an octave-spanning frequency comb compared to the same resonator with a straight waveguide coupling.

\section*{Acknowledgement}
This work is supported by the DARPA DODOS, ACES, and NIST-on-a-chip programs. G.M., X.L., Q.L. and A.R. acknowledge support under the Cooperative Research Agreement between UMD and NIST-PML, Award no. 70NANB10H193.
\bibliography{Coupling}
% \bibliographyfullrefs{Coupling}
\end{document}